\documentclass{ws-ijmpe}
\usepackage{citesort}
\begin{document}

\markboth{Monika Sharma, Sunil Dogra, Neeraj Gupta}{Energy And System Size Dependence of Photon Production at Forward Rapidities at RHIC}

\catchline{}{}{}{}{}

\title{\bf ENERGY AND SYSTEM SIZE DEPENDENCE OF PHOTON PRODUCTION AT FORWARD RAPIDITIES AT RHIC}

\author{MONIKA SHARMA\scriptsize$^{1}$, SUNIL DOGRA\scriptsize$^{2}$, NEERAJ GUPTA\scriptsize$^{2}$\\
(for the STAR Collaboration)}
\address{\scriptsize$^{1}$Department of Physics, Panjab University, Chandigarh - 160014, India\\
monika@rcf.rhic.bnl.gov}
\address{\scriptsize$^{2}$Department of Physics, University of Jammu, Jammu, India\\
sunil@rcf.rhic.bnl.gov, neeraj@rcf.rhic.bnl.gov}

\maketitle

\begin{history}
\received{(received date)}
\revised{(revised date)}
\end{history}

\begin{abstract}
The energy and system size dependence
of pseudorapidity ($\eta$) and multiplicity distributions of photons
 are measured in the region $-2.3$ $\leq$ $\eta$ $\leq$ $-3.7$ for Cu + Cu
collisions at $\sqrt{s_{NN}}$ = 200 and 62.4 GeV. Photon multiplicity measurements at forward rapidity
have been carried out using a Photon Multiplicity Detector (PMD)
in the STAR experiment. Photons are found to follow longitudinal scaling for Cu + Cu collisions for
0-10$\%$ centrality. A Comparison of pseudorapidity distributions with
HIIJING model is also presented.\\
\end{abstract}

\section{Introduction}
One of the major goals of Relativistic Heavy Ion Collider at Brookhaven National Laboratory, Upton
(NewYork), is to search for the possible formation of Quark Gluon Plasma in heavy ion 
collisions\cite{qcd1}. Multiplicity and pseudorapidity distributions are one of
the very first
 measurements made at RHIC. 
One can obtain important information about the collision
dynamics by studying dependences of the particle multiplicity and pseudorapidity
distributions on collision centrality, energy, system size etc.
Multiplicity
distributions have been used to understand the particle production mechanism
based on participant scaling, binary scaling, two component model\cite{tcm} and recently
by invoking the Color Glass Condensate (CGC)\cite{cgc} model. Pseudorapidity distributions
coupled with the measurement of average transverse energy provide information about the 
energy density achieved in the collision using the Bjorken formula\cite{BF} and
on the nature of the system produced using hydrodynamics with CGC\cite{cgc} as the initial condition.\\
A lot of work has been reported on measurements of the charged particles produced in heavy ion collisions
covering complete pseudorapidity region\cite{char1} \cite{char2}.
But small amount of work is available for photons produced in such collisions in forward rapidity region.
Only preshower detectors at Super Proton
Synchrotron (SPS) and the STAR at RHIC have explored this
region of pseudorapidity\cite{ch2}-\cite{pho2}.\\  
Photons are considered as one of the most valuable probes of the dynamics
and properties
of the matter formed in the heavy ion collisions as they interact only
electromagnetically\cite{pho2}. Photons have a large mean free path and hence carry
the first hand information of their origin.
There are predictions of more direct photon production specifically associated to QGP
formation in the
heavy ion collisions\cite{dp}. However, the main contribution comes from the decay of
$\pi^{0}$'s
 produced in the collisions during hadronization. Pseudorapidity distributions are used in
 validating the theoretical 
models attempting to describe the conditions in the early state of the collision\cite{model} \cite{wa98} \cite{pmd}.\\
It has been found that at forward rapidity regions, charged
particle pseudorapidity distributions show a longitudinal scaling. The variation of rapidity
density per participant pair with ($\eta$ - Y$_{beam}$) where $\eta$ is the pseudorapidity and
Y$_{beam}$ is the beam rapidity, is found to be independent of energy. Centrality dependence of
 such a behaviour has been studied by BRAHMS\cite{char1}, PHOBOS\cite{char2} and STAR\cite{ch2}.
The STAR experiment reported measurements of the
pseudorapidity distribution in the forward rapidity region ($-2.3 \leq \eta \leq -3.7$)
in Au + Au collisions at $\sqrt{s_{NN}}$ = 62.4 GeV using the preshower Photon Multiplicity Detector 
(PMD)\cite{pmd}. The photon yield scales with the number of participating nucleons and follow 
longitudinal pseudorapidity scaling away from the mid-rapidity which is independent of energy.
Limiting Fragmentation (LF) hypothesis\cite{LF} is used to
explain this, but recently CGC\cite{cgc} is also used to understand the effect at forward rapidities. 
B.B. Back {\it et al.}\cite{ch}, 
observed that pseudorapidity distributions of charged particles for central Au + Au and Cu + Cu collisions
 exhibit the same shape over six units of rapidity (i.e., $|\eta|<3$). 
The ratios of $dN_{ch}/d\eta$ between Au + Au and Cu + Cu 
collisions are constant at a value of 3.6 for 62.4 GeV and 3.56 for 200 GeV and are
 slightly more than the ratio of number of nucleons available in the initial state i.e., 
$A_{Au}/A_{Cu}$ = 3.13. For $|\eta|>3$ they fall off more steeply 
for 62.4 GeV than for 200 GeV.\\ 
In view of the above work, it is important to investigate the energy as well as
system size dependence of pseudorapidity distributions of photons. In this paper,
we present the pseudorapidity distributions of photons for 
 Cu + Cu collisions at $\sqrt{s_{NN}}$ = 200 GeV and 62.4 GeV.
These studies have been carried out for different collision centralities. Results
have been compared with HIJING monte carlo event generator\cite{hijing}.

\section{Experimental Setup and Data Analysis}
\begin{figure}[h]
  \hfill
  \begin{minipage}[t]{.45\textwidth}
    \begin{center}
      \centerline{\resizebox{7.0cm}{5.0cm}{\epsfig{file=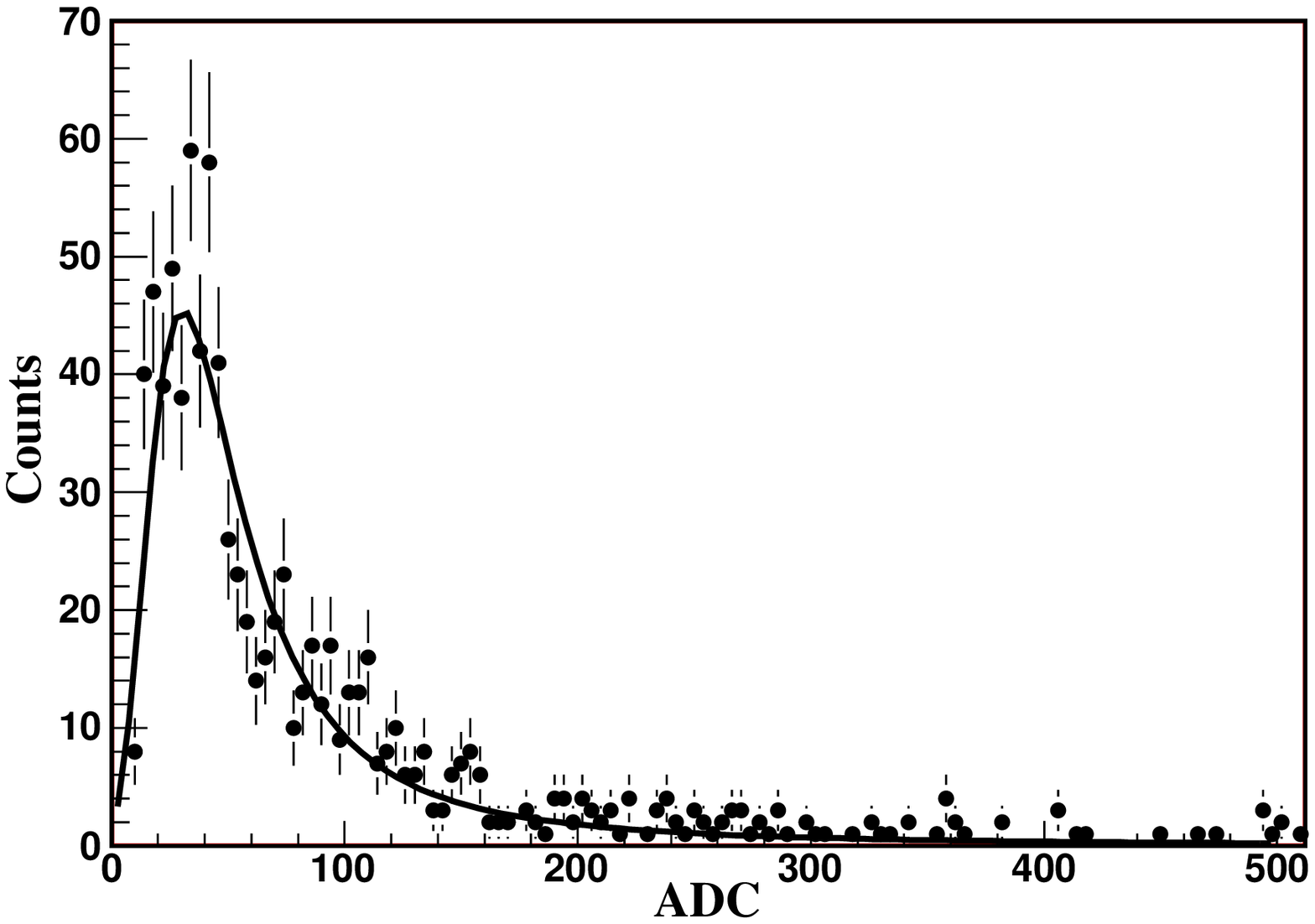, scale=0.45}}}
      \caption{ADC distribution of an isolated cell.}
      \label{fig1}
    \end{center}
  \end{minipage}
  \hfill
  \begin{minipage}[t]{.45\textwidth}
    \begin{center}
      \centerline{\resizebox{7.0cm}{5.0cm}{\epsfig{file=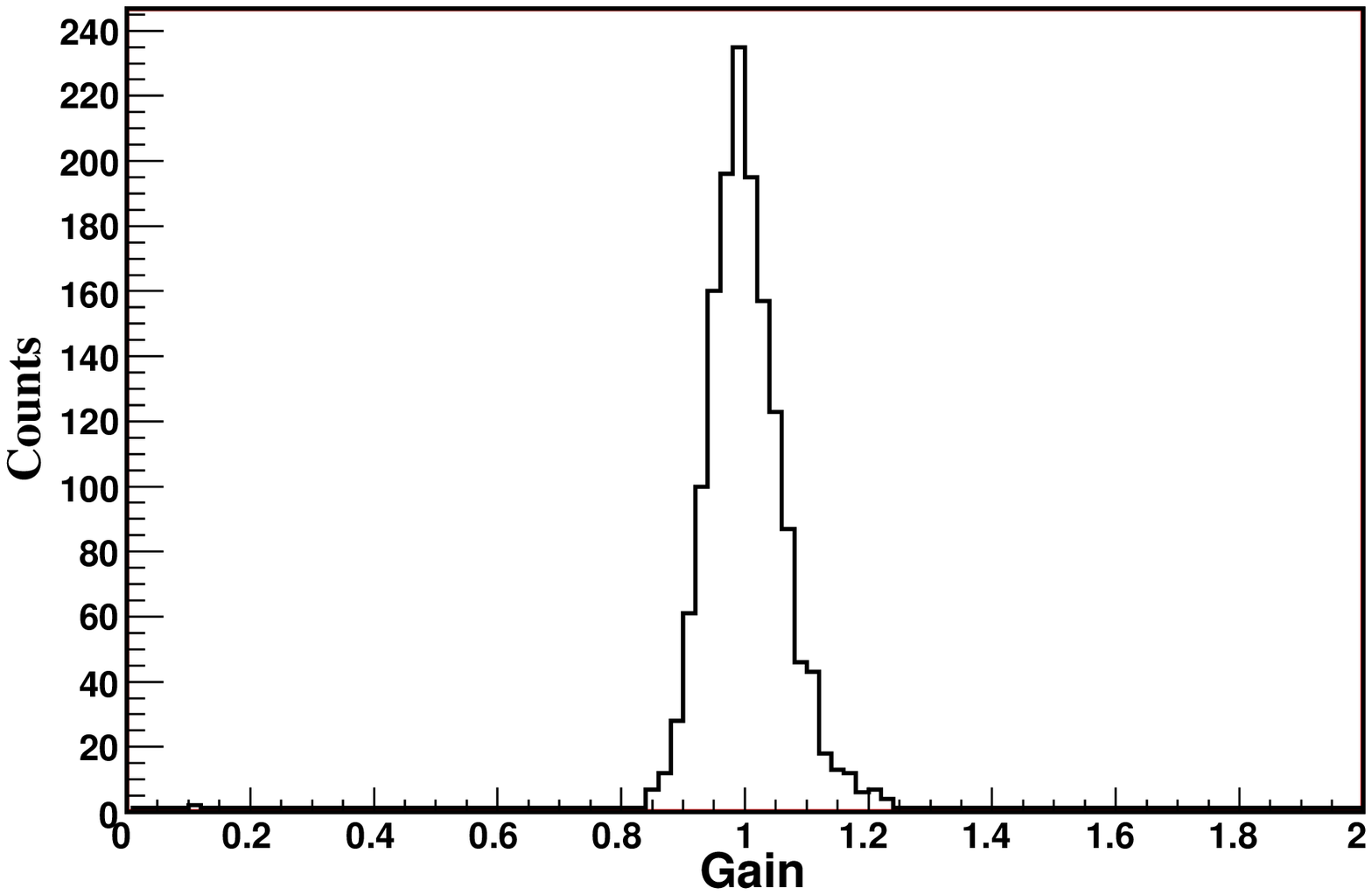, scale=0.45}}}
      \caption{Relative cell-to-cell gain distribution.}
      \label{fig2}
    \end{center}
  \end{minipage}
  \hfill
\end{figure}

The PMD is placed at a Z-distance of 5.4 m from the center of TPC
(the nominal collision point) along the beam axis. It consists of a highly
segmented gaseous detector on a plane placed behind a lead converter plate of 3 radiation 
length (3$X_{0}$) thick\cite{PMD}, known as preshower plane. A veto plane which is also a 
gaseous detector is placed in front of the converter to reject the charged particles. 
The planes are further sub-divided into 12 gas tight entities, known as
supermodules (SMs). Veto plane is not used in this analysis.
Discrimination between photons and charged hadrons is done by their difference in response e.g.,  
charged hadrons affect mostly one cell with Minimum Ionising Particle (MIP) like energy deposition, whereas 
the number of cells affected and signal from photon are large. \\
Uniformity of the detector is obtained by finding MIP-response from each cell.
MIP response of each cell is obtained by selecting cells having a signal surrounded by
six cells without any signal representing an isolated cell.
Fig. \ref{fig1} displays the ADC distribution of an isolated cell which follows
a Landau distribution with a mean of 84.61 ADC and most probable value (MPV) of
33.41 ADC. The relative gain for each cell is computed by dividing the cell ADC mean by the average mean of all
cells in a SM.
 Fig. \ref{fig2} displays a typical cell-to-cell relative gain distribution for one SM.
 Photons in an event are counted by finding clusters from cells with non-zero signal and 
applying a suitable cut on the cluster signal and number of cells to reject charged hadrons.
Following criteria is evolved to find photon like 
clusters ($N_{\gamma-like}$)  using the HIJING Monte Carlo event generator
 + GEANT\cite{geant} : (a) the number of hit cells in a cluster $>$ 1 and (b) the cluster 
signal is 3 times or more than the average response of all isolated cells in
a SM. Similar threshold is also applied in data to count number of photon like clusters event-by-event.
The number of photons (N${_\gamma}$) from the ($N_{\gamma-like}$) are obtained as :
\begin{equation}
N_{\gamma} = N_{\gamma-like}*\frac{f_{p}}{\in_{\gamma}}\\
\label{data}
\end{equation}
Simulations have been performed by running full GEANT with STAR geometry (GSTAR) using 
HIJING events for obtaining efficiency ($\in_{\gamma}$) and purity ($f_{p}$).  
Photon reconstruction efficiency  ($\in_{\gamma}$) and purity ($f_{p}$) are calculated as :
\begin{equation}
\in_{\gamma} = \frac{N^{\gamma,th}_{cls}}{N_{inc}^{\gamma}}\\
\end{equation} 
\begin{equation}
f_{p} = \frac{N_{cls}^{\gamma,th}}{N_{\gamma-like}}
\end{equation} 
Here $N_{cls}^{\gamma,th}$ is the number of photon clusters identified above the hadron rejection threshold
 and $N_{inc}^{\gamma}$ are the number of incident photons. Both $N_{cls}^{\gamma,th}$ and $N_{inc}^{\gamma}$ are 
obtained from the event generator. The geometrical acceptance factors defined below are obtained pseudorapidity 
binwise for the SMs used in the present PMD analysis. \\
\begin{equation}
A = \frac{Total\hspace{0.2cm}cells\hspace{0.2cm}within\hspace{0.2cm}the\hspace{0.2cm}pseudorapidity\hspace{0.2cm}bin}{Total\hspace{0.2cm}number\hspace{0.2cm}of\hspace{0.2cm}active\hspace{0.2cm}cells}
\end{equation}
For the present analysis, we have selected SMs with stable gain throughout the data taking. Also, cells with abnormally high frequency of hits were treated as dead cells.
In order to implement the SM to SM gain
variation, we have calculated average MIP for each SM from data. Taking the SM with best developed
 MIP as standard, the variation of SM-wise gain has been incorporated in simulations and the responses 
of cells are changed accordingly.\\
\section{Results and Discussions}
\begin{figure}[h]
  \hfill
  \begin{minipage}[t]{.45\textwidth}
    \begin{center}  
      \centerline{\resizebox{7.0cm}{5.0cm}{\epsfig{file=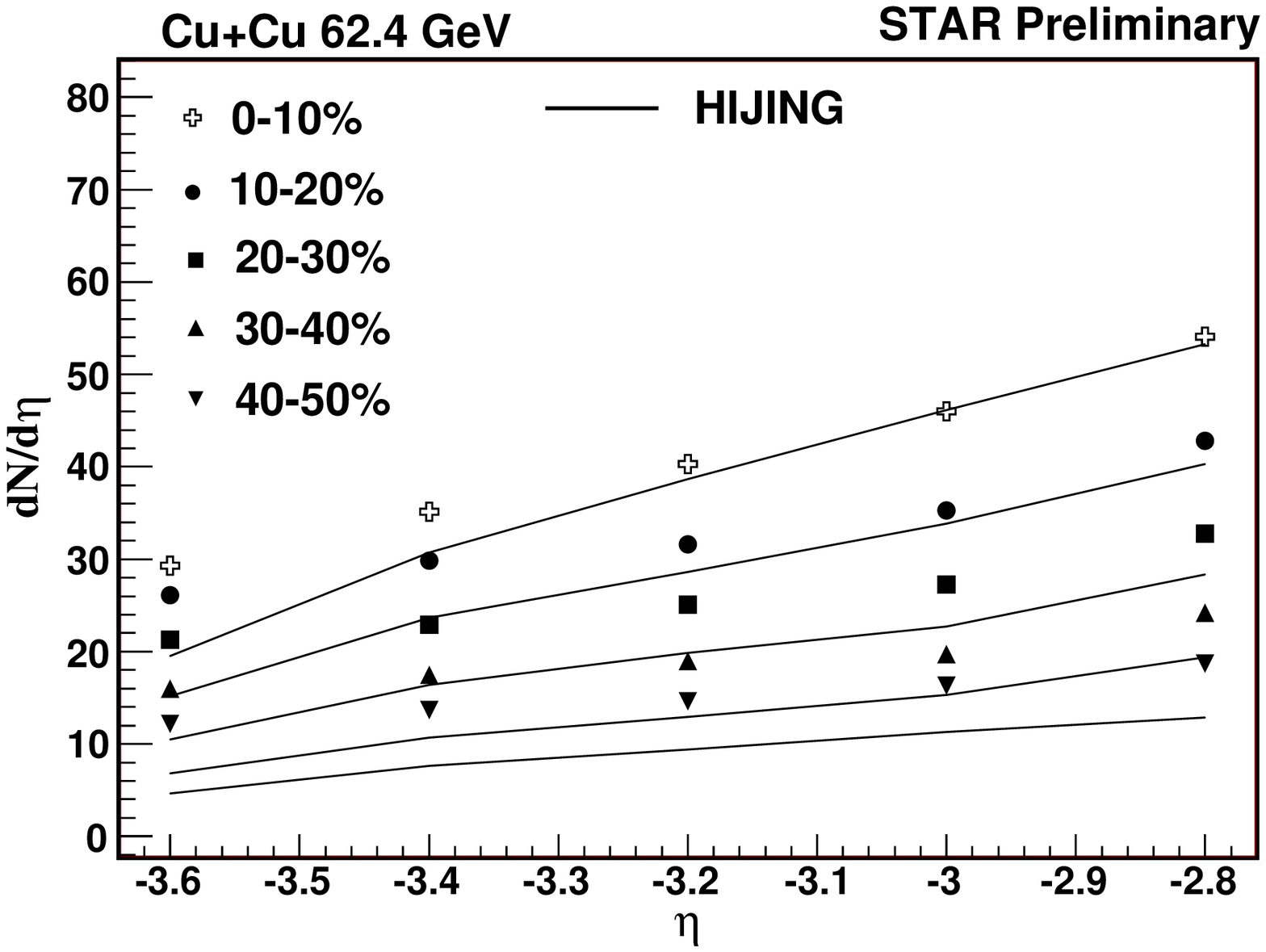, scale=0.45}}}
      \caption{Photons pseudorapidity density distribution, $\frac{dN_{\gamma}}{d\eta}$, measured for Cu + Cu at $\sqrt{s_{NN}}$ = 62.4 GeV. The statistical errors are within the symbol size.}
      \label{mm}
    \end{center}
  \end{minipage}
  \hfill
  \begin{minipage}[t]{.45\textwidth}
    \begin{center}  
      \centerline{\resizebox{7.0cm}{5.0cm}{\epsfig{file=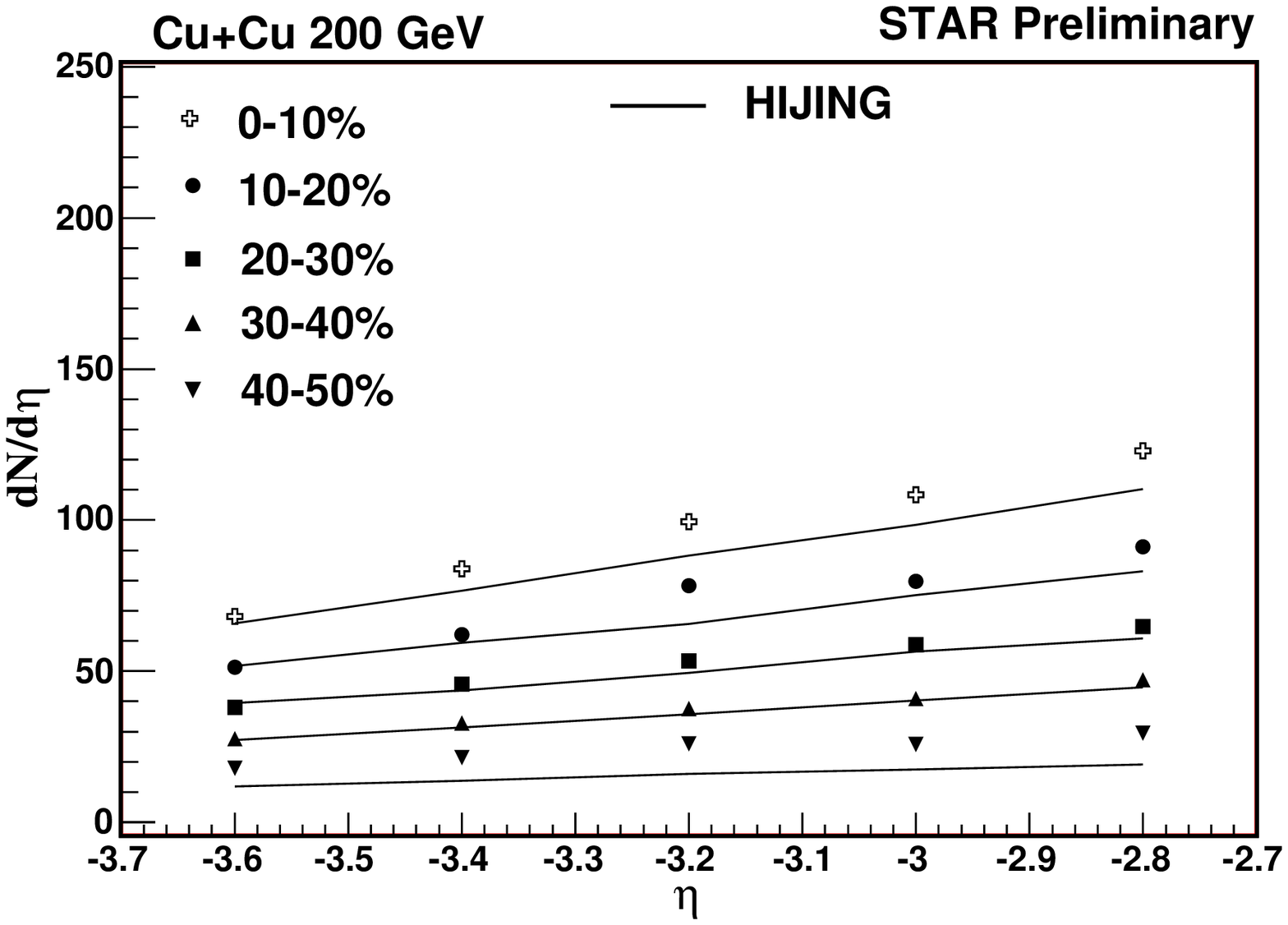, scale=0.45}}}
      \caption{Photons pseudorapidity density distribution, $\frac{dN_{\gamma}}{d\eta}$, measured for Cu + Cu at $\sqrt{s_{NN}}$ = 200 GeV. The statistical errors are within the symbol size.}
      \label{mon1}
    \end{center}
  \end{minipage}
  \hfill
\end{figure}

\begin{figure}[th]
\centerline{\resizebox{10.0cm}{5.0cm}{\psfig{file=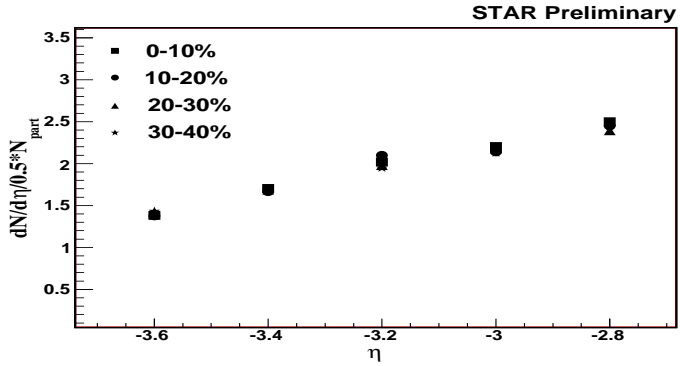,width=10cm}}}
\vspace*{8pt}
\caption{Photons pseudorapidity distribution per participant pair, $\frac{\frac{dN_{\gamma}}{d\eta}}{0.5*N_{part}}$, measured for Cu + Cu at $\sqrt{s_{NN}}$ = 200 GeV for different centralities. The statistical errors are within the symbol size.}
\label{mon}
\end{figure}

The data for Cu + Cu collisions at 
$\sqrt{s_{NN}}$ = 200 and 62.4 GeV, taken during the year 2004 $\&$ 2005 are presented in this paper.
Only minimum bias events are taken for this analysis which was obtained by coincidence between two 
Zero Degree Calorimeters (ZDCs) and a minimum signal from Central Trigger Barrel (CTB). Events which were 
produced within $\pm$50 cm of the center of the TPC along the beam axis were accepted for analysis. 
The centrality determination of this analysis uses the multiplicity of charged particles in the pseudorapidity 
range $|\eta|<$0.5, as measured by TPC.\\
Fig. \ref{mm} shows the measured pseudorapidity distributions for photons for several centrality classes in
 Cu + Cu collisions at $\sqrt{s_{NN}}$ = 62.4 GeV. For comparison HIJING predictions are also displayed in 
the figure. Fig. \ref{mon1} exhibits the pseudorapidity distributions for photons for
different centrality in Cu + Cu collisions at $\sqrt{s_{NN}}$ = 200 GeV. 
 It is seen that matching with HIJING is better towards the central events.
\\
Fig. \ref{mon} shows $\frac{dN_{\gamma}}{d\eta}$, scaled by the number of participating 
nucleons for each centrality. It is observed that $\frac{dN_{\gamma}}{d\eta}$, scaled by the number of participating 
nucleons, is independent of centrality.
\begin{figure}[th]
\centerline{\resizebox{10.0cm}{6.0cm}{\psfig{file=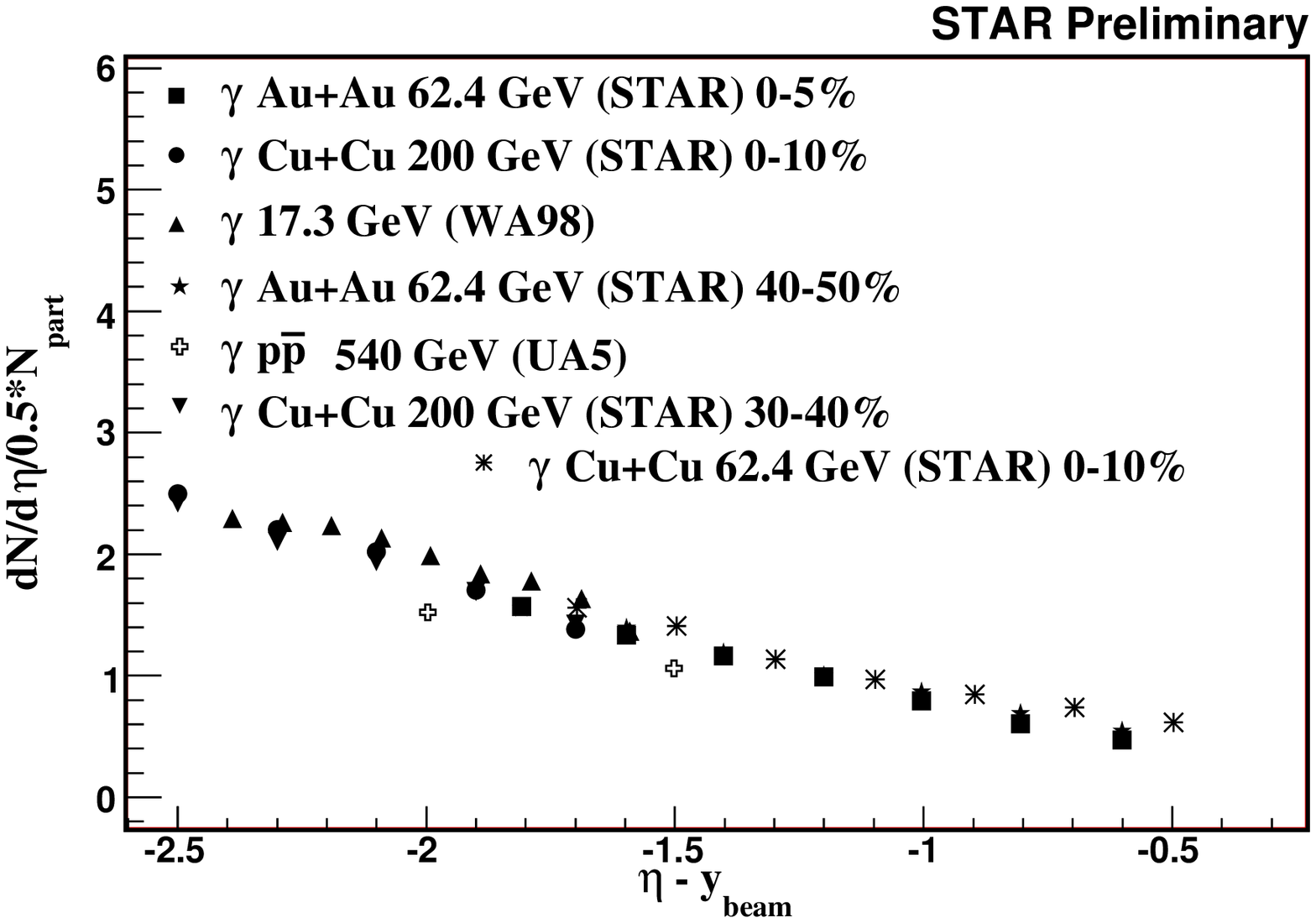,width=10cm}}}
\vspace*{8pt}
\caption{Photons pseudorapidity distribution per participant pair, $\frac{\frac{dN_{\gamma}}{d\eta}}{0.5*N_{part}}$ as a function of $\eta-y_{beam}$ for different energies $\&$ systems as indicated. The statistical errors are within the symbol size.}
\label{mon2}
\end{figure}
Fig. \ref{mon2} shows the longitudinal scaling for photons at different energies
and centralities. Here, we compare pseudorapidity distribution per participant
pair for Au + Au central  (0-5\%) and peripheral (40-50\%) events at
$\sqrt{s_{NN}}$ = 62.4 GeV, for  Cu + Cu central (0-10\%) and peripheral 
(30-40\%) events at $\sqrt{s_{NN}}$ = 200 GeV as a function of $\eta-y_{beam}$. The WA98\cite{wa98} data
at $\sqrt{s_{NN}} = 17.3$ GeV and the UA5\cite{ua5} data for p$\bar{p}$ at $\sqrt{s_{NN}}$ = 540 GeV are
also displayed. We observe that photons follow universal limiting pseudorapidity distribution away 
from mid rapidity which is independent of energy, centrality and system size.
\section{Summary}
The measurements of the pseudorapidity distributions from Cu + Cu collisions
at top RHIC energy ($\sqrt{s_{NN}}$ = 200 GeV) and 62.4 GeV have been
presented. Results have been compared with the HIJING Model. Photons pseudorapidity 
distributions follow limiting longitudinal scaling  
away from the mid rapidity. 
We further observe that longitudinal scaling is not only independent of energy but also independent of centrality and system size.

\end{document}